\pdfoutput=1
\documentclass[aip,jap,reprint,superscriptaddress,floatfix]{revtex4-1}

\usepackage[]{graphicx}
\usepackage{amssymb}
\usepackage{epstopdf}
\usepackage{soul}
\usepackage{xcolor}
\usepackage{amsmath}
\usepackage{gensymb}
\usepackage[nolist]{acronym}
\usepackage{bm}
\def\micron{$\mu$m}

\def\Lone{\textit{L1}}
\def\Ltwo{\textit{L2}}
\def\D{\textit{D}}
\def\F{\textit{F1}}
\def\Fone{\textit{F1}}
\def\Ftwo{\textit{F2}}

\def\mathrelfun#1#2{\lower3.6pt\vbox{\baselineskip0pt\lineskip.9pt
  \ialign{$\mathsurround=0pt#1\hfil##\hfil$\crcr#2\crcr\sim\crcr}}}

\newcommand\apjs{{Ap.~J.~Suppl.~}}

\newcommand\procspie{{\it Proceedings of the SPIE}}
  
\newcommand\pasj{{PASJ}}               % Publications of the ASJ

\newcommand\ben{\begin{eqnarray}}
\newcommand\een{\end{eqnarray}}

\begin{document}

%\preprint{}

\title{Broadband Millimeter-Wave Anti-Reflection Coatings on Silicon Using Pyramidal Sub-Wavelength Structures} 

\author{Karl Young}
\email[Electronic mail: ]{kyoung@astro.umn.edu}
\affiliation{School of Physics and Astronomy, and Minnesota Institute for Astrophysics, University of  Minnesota/Twin Cities, 116 Church St. SE Minneapolis, MN 55455, USA}

\author{Qi Wen}
\affiliation{School of Physics and Astronomy, and Minnesota Institute for Astrophysics, University of  Minnesota/Twin Cities, 116 Church St. SE Minneapolis, MN 55455, USA}

\author{Shaul Hanany}
\affiliation{School of Physics and Astronomy, and Minnesota Institute for Astrophysics, University of  Minnesota/Twin Cities, 116 Church St. SE Minneapolis, MN 55455, USA}

\author{Hiroaki Imada}
\affiliation{Japan Aerospace Exploration Agency (JAXA) - Institute of Space and Astronautical Science (ISAS), 3-1-1 Yoshinodai, Chuo, Sagamihara, Kanagawa 252-5210, Japan.}

\author{J\"{u}rgen Koch}
\affiliation{Laser Zentrum Hannover e.V., Hollerithallee 8, 30419 Hannover, Germany}

\author{Tomotake Matsumura}
\affiliation{Kavli Institute for the Physics and Mathematics of the Universe (Kavli IPMU, WPI), Todai Institutes for Advanced Study, The University of Tokyo, 5-1-5 Kashiwa-no-Ha, Kashiwa City, Chiba 277-8583, Japan}

\author{Oliver Suttmann}
\affiliation{Laser Zentrum Hannover e.V., Hollerithallee 8, 30419 Hannover, Germany}

\author{Viktor Sch\"{u}tz}
\affiliation{Laser Zentrum Hannover e.V., Hollerithallee 8, 30419 Hannover, Germany}

\date{\today}

\begin{abstract}

We used two novel approaches to produce sub-wavelength structure (SWS) anti-reflection coatings (ARC) on silicon 
for the millimeter and sub-millimeter (MSM) 
wave band: picosecond laser ablation and 
dicing with beveled saws.  We produced pyramidal structures with both techniques. The diced 
sample, machined on only one side, had pitch and height of 350~\micron\ and 972~\micron. The 
two laser ablated samples had 
pitch of 180~\micron\ and heights of 720~\micron\ and 580~\micron; only one of these samples was ablated on both sides. 
We present measurements of shape and optical performance
as well as comparisons to the optical performance predicted using finite element analysis and rigorous coupled
wave analysis. 
By extending the measured performance of the one-sided diced sample to the two-sided case, we demonstrate 25~\% 
band averaged reflectance of less than 5~\% over a bandwidth 
of 97~\% centered on 170~GHz.  
Using the two-sided laser ablation sample, we demonstrate reflectance less than 5~\% over 83~\% 
bandwidth centered on 346~GHz.

\end{abstract}


\maketitle %\maketitle must follow title, authors, abstract and \pacs

\section{Introduction}

Silicon is an appealing optical material for the \ac{MSM} region of the electromagnetic spectrum, 
approximately between 30 and 3000~GHz. It has a high index of refraction $n=3.4$ and low 
loss\cite{lamb96} $\tan\delta < 10^{-4}$. It thus gives higher aberration correction 
power and higher transmission efficiency compared to plastic lenses, as well as easier machinability and lower 
loss compared to alumina, which has a similar index of refraction.  
The high index of refraction also causes substantial reflections. Without \ac{ARC} the two surfaces
of a 10~mm thick disc give band-averaged ($\Delta \nu / \nu = 30~\%$) reflectance of 46~\%. 

Silicon is used as a lens material for astrophysical instruments in the 
\ac{MSM}\cite{thornton2016_actpol,piper_optics, mitsui2015_silicon_lenslet,edwards2012_silicon_lenslet,sekiguchi2015}
and as a lens and grism material in the infrared.\cite{akari2007, forcast_grism, nircam_jwst}
Various \ac{ARC} approaches have been implemented including gluing Cirlex,\cite{lau2006_cirlex} vapor 
deposition of Parylene,\cite{gatesman2000_paryleneC,wheeler2014} lithography and ion etching of 
\ac{SWS},\cite{wheeler2014,kamizuka2012} and \ac{SWS} cut using standard dicing saws.\cite{datta2013} 
Fabricating \ac{SWS} is a particularly appealing \ac{ARC} technique because it can 
provide a broadband coating without the need to match indices of several materials 
and because it is robust to cryogenic cycles.

In this paper we present two novel approaches to fabricating \ac{SWS}-\ac{ARC} on silicon: laser ablation 
and dicing with beveled saws. Our ultimate motivation is the development of scalable techniques to produce
lenses for broadband \ac{CMB} polarization instruments
which are observing with fractional bandwidths of 60--110\%\cite{spt-3g,spt3g_tes_bands, thornton2016_actpol,polarbear2_inst} 
and are planned with 150\% fractional bandwidth.\cite{Matsumura2016_LB}
With these instruments the lenses 
are typically maintained at cryogenic temperatures, and it is important they exhibit low instrumental polarization. 

We have recently demonstrated the first laser ablated \ac{SWS}-\ac{ARC} on sapphire and alumina.\cite{schutz2016, matsumura2016_arc} 
Ablation of silicon has been investigated under different conditions,\cite{Phillips2015_review} 
including varying power levels,\cite{domke2016,Bonse2002} 
wavelengths,\cite{schutz2013thermodynamic} 
and pulse durations.\cite{her1998,zhao2003}
Several authors investigated the use of gas-assisted laser ablation to produce \ac{SWS} on silicon.\cite{younkin2003,sheely2005}
The grid spacing of the resulting structures makes them most suitable for visible and near infrared 
wavelengths. To our knowledge, 
this paper is the first to report on the ablation of \ac{SWS}-\ac{ARC} for the \ac{MSM} waveband. 

In Section~\ref{sec:fab} we describe our samples and the \ac{SWS} fabrication. 
In Section~\ref{sec:measurements} we describe and discuss measurements of the shapes of the \ac{SWS}.
We discuss transmission and reflection measurements between 70 and 700~GHz in Section~\ref{sec:reflect},
and summarize our results in Section~\ref{sec:summary}.

\begin{table*}[htbp]
   \centering
   \caption{Physical properties of each sample.
   \label{tab:samplelist}}
   \begin{tabular}{c|c|c|c|c|c|c} 
   \hline \hline
	    Sample & Method & Sides Coated & Refractive Index & Thickness & Diameter & Resistivity  \\ 
	 \ & \  & \ & \ &  [mm] & [mm] & [$\ohm\cdot$cm]  \\ \hline
	Flat1 (\textit{F1})& No ARC & No ARC & 3.405 $\pm$ 0.002 & 2.002 $\pm$ 0.001 & 50.8 & $>1000$ \\
	Flat2 (\textit{F2})& No ARC & No ARC & 3.417 $\pm$ 0.002 & 6.013 $\pm$ 0.002 & 50.8 & $>500$ \\
	Laser1 (\textit{L1}) & Laser & one side & 3.405 $\pm$ 0.002 & 2.009 $\pm$ 0.001 & 50.8 & $>1000$ \\
	Laser2 (\textit{L2}) & Laser & both sides & 3.417 $\pm$ 0.002  & 6.009 $\pm$ 0.002 & 50.8 & $>500$ \\ 
	Diced (\textit{D}) & Dicing saw & one side & 3.405 $\pm$ 0.002 & 2.009 $\pm$ 0.001 & 50.8 & $>1000$ \\ 
   \hline \hline
   \end{tabular}
\end{table*}

\section{Samples and Fabrication}
\label{sec:fab}

\subsection{Samples} 

We fabricated two samples using laser machining, called \Lone\ and \Ltwo, 
and one using a dicing saw, called \D. \Lone\ and \D\ were processed on one\ side only; \Ltwo\ was processed 
on both sides.  We also measured two unprocessed flat discs, called \Fone\ and \Ftwo. These samples were  
used to cross-check the measurements against analytic predictions and to determine the index 
of refraction and loss tangents of the other samples. 

\begin{figure}[h] 
   \centering
   \includegraphics[width=8.5cm]{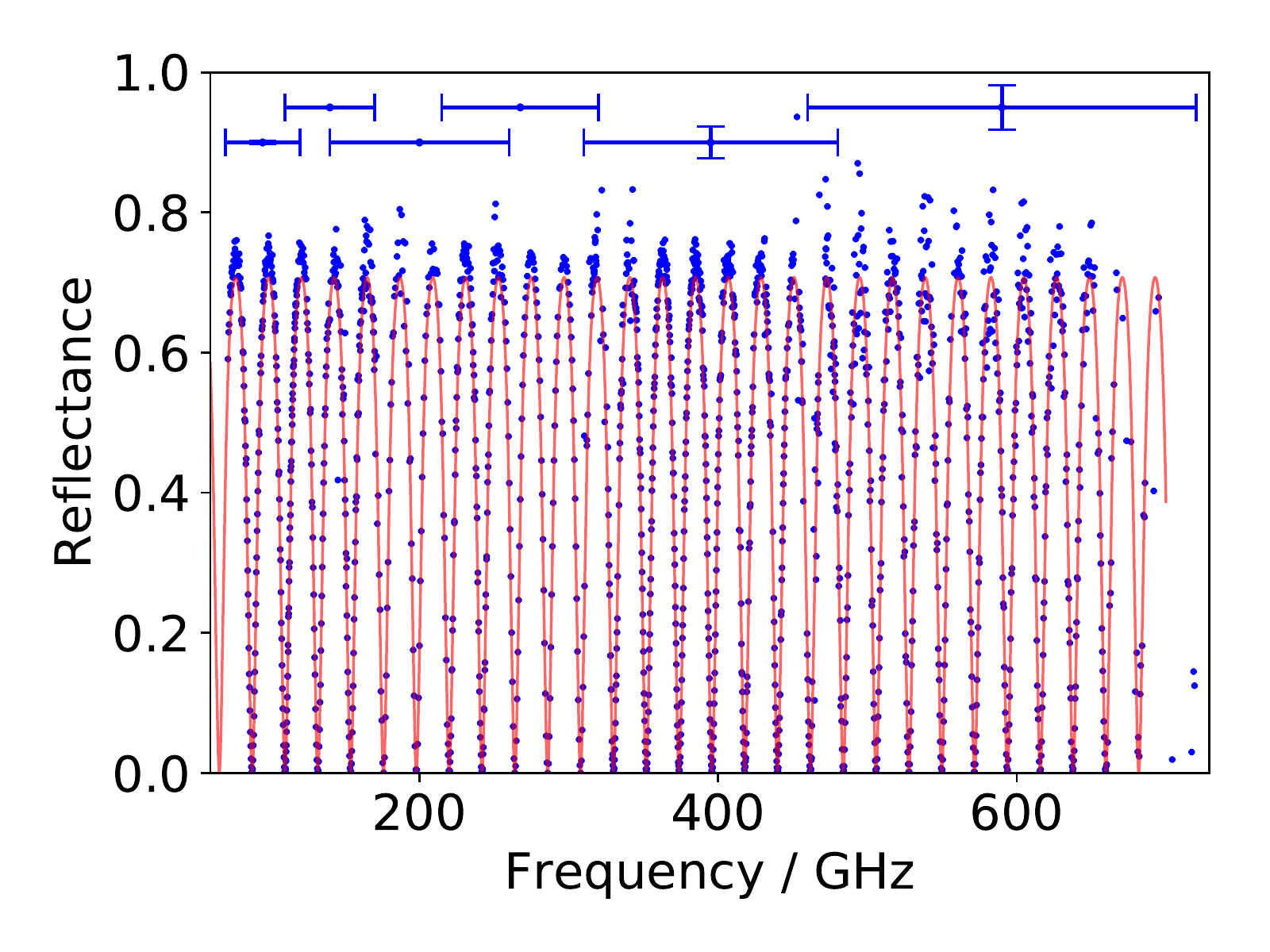} 
   \caption{ Reflectance of a flat 2~mm thick silicon disc, sample \Fone\ (blue points), and 
    theoretical prediction (red, solid) with the best fit
    index $n = 3.405$ and no loss. 
    Measurement errors (blue) are discussed in Section~\ref{sec:measurementprocedure}.
   \label{fig:silicon_flat_ucsb} }
\end{figure}

The high-resistivity silicon discs for \Fone , \D ,  and \Lone\ were part of the same 
order\footnote{Miyoshi Ltd. \protect\url{park15.wakwak.com/~miyoshi}, Japan} and arrived in the 
same shipment. We therefore assumed the same material properties for all three.  
Using reflectance measurements we fit for the index and loss tangent of \Fone\ and found $n=3.405$ 
and loss tangent $<10^{-4}$; see Figure~\ref{fig:silicon_flat_ucsb}. 
Samples \Ftwo\ and \Ltwo\ were 
from a second order and shipment, \footnote{University Wafer Inc. \protect\url{universitywafer.com}, USA} 
so we assumed the same material properties for both.
We measured transmittance of \Ftwo, fit for index and loss, and 
found $n=3.417$ and loss tangent $<10^{-4}$. 
Table~\ref{tab:samplelist} summarizes the information about the samples. 

\subsection{Laser ablation}

\begin{figure}[h] 
   \centering
   \includegraphics[width=8.5cm]{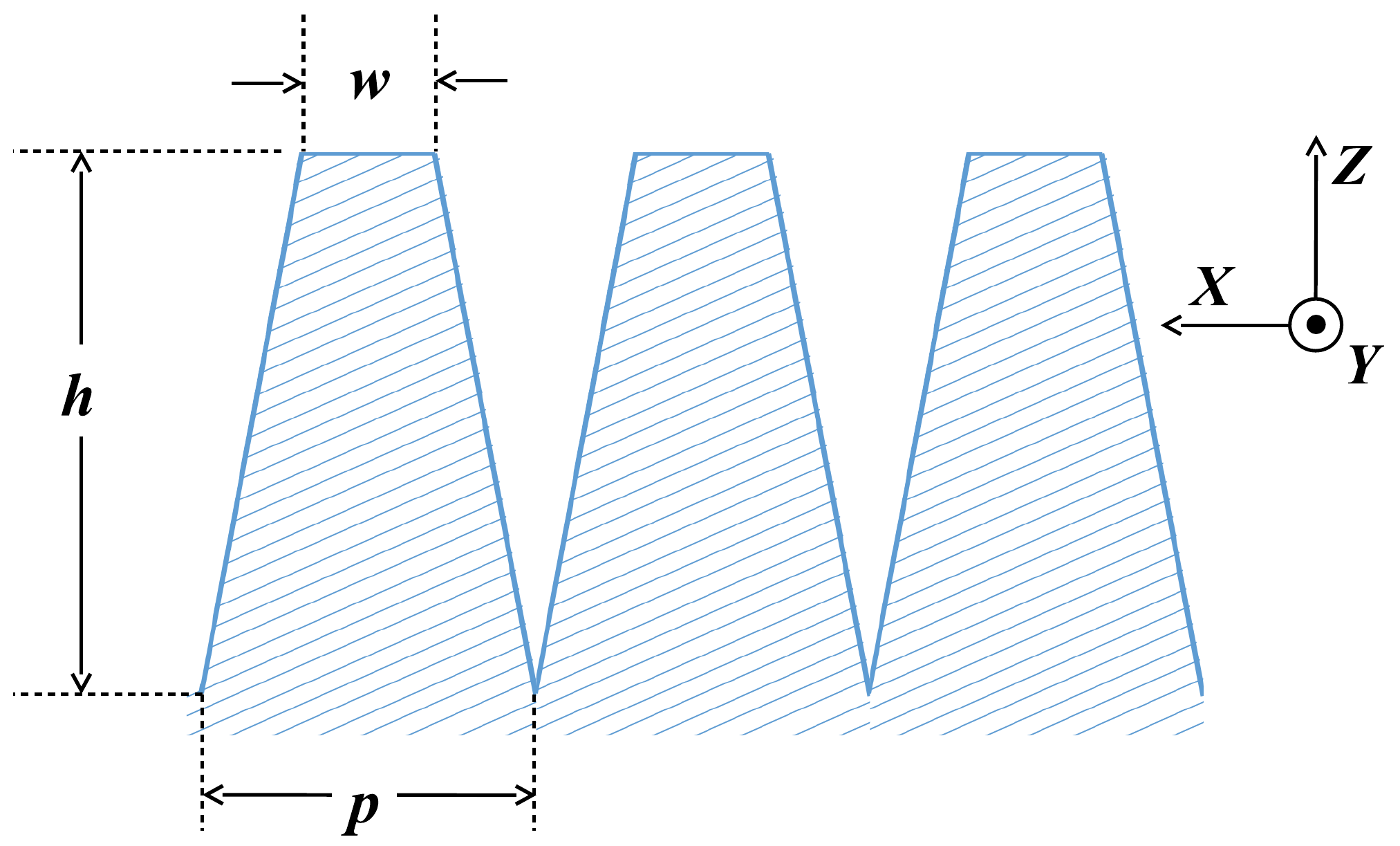} 
   \caption{ Side view sketch defining key shape parameters of the \ac{SWS}. The hatched region
   is silicon.  The changing amount of silicon relative to vacuum creates a 
            gradual change in index of refraction along the $z$-axis.
   \label{fig:size} }
\end{figure}

\begin{figure}[h] 
   \centering
   \includegraphics[width=8.5cm]{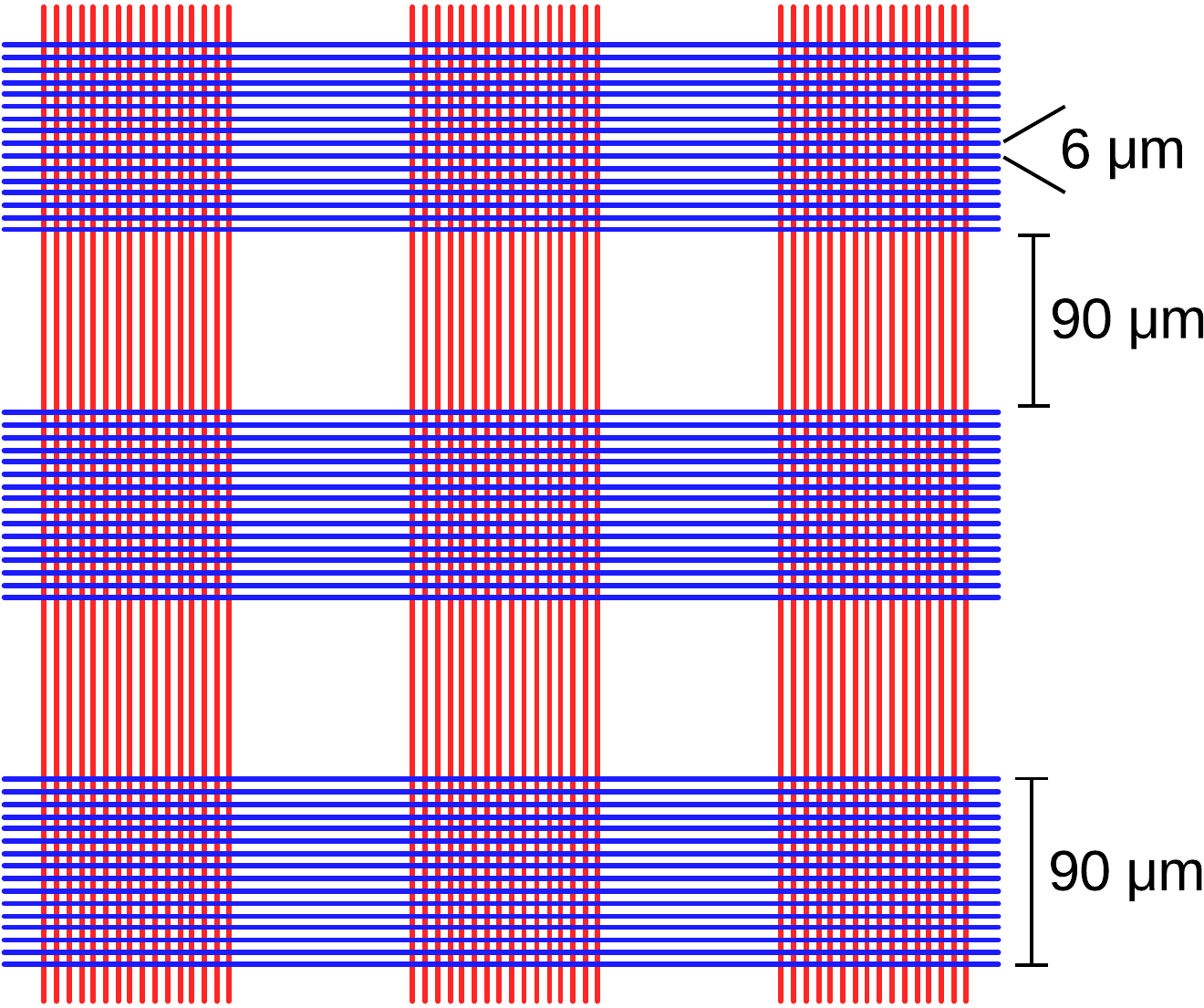} 
   \caption{The laser scanned \Lone\ and \Ltwo\ in a raster pattern.  A single `scan' consists of sequentially 
   passing through 
   all horizontal (blue) lines, then all vertical (red) lines. The scans were repeated 80 times for both \Lone\ and \Ltwo.  
   Each group of blue lines spaced by 6~\micron\ eventually makes a single groove.  The 90~\micron\ gaps are 
   the areas where material is not ablated, leaving the tips of the pyramids. 
   \label{fig:scan} }
\end{figure}

We machined \Lone\ and \Ltwo\ with parameters that were similar to our earlier laser ablation of \ac{SWS} 
on alumina and sapphire;\cite{matsumura2016_arc}  the parameters 
of the geometry and the coordinate system are shown in Figure~\ref{fig:size}.
The laser operated at a wavelength of 515~nm 
with a repetition rate of 400~kHz, 7~ps pulse width, 
and an average power of 28.5~W for \Lone\ and 27~W for \Ltwo. 
The beam was focused at the surface of the silicon disc and had a $1/e^2$ width of 8~\micron.  
The laser was scanned across the entire sample 
at 2.5 m/sec in a raster pattern as shown in Figure~\ref{fig:scan}.  This pattern created 
a series of orthogonal grooves and pyramids with a designed pitch of 180~\micron.  
The scan was repeated 80 times. 
The total machining time for \Lone\ was 3.4 hours for a 5~cm diameter disc.
For \Ltwo\ the grid patterns on the two sides were oriented at 45\degree\ with respect to each other. 
The total machining time for both sides was 6.8 hours.  We did not attempt to optimize the ablation rate.

\subsection{Dicing saw}

Sample \D\ was machined using a custom made beveled dicing saw. 
The blade had a maximum thickness of 300~\micron. Imaging the profile of the first cut,
we measured that the blade tip was $\leq 10$~\micron\ wide and the bevel angle was 
81.6\degree\ providing a usable cutting depth of 1.04~mm. 
We produced pyramids by cutting 108 parallel grooves symmetrically placed about a diameter 
with a feed rate of 1~mm/sec, rotating the sample by $90\degree \pm 0.05$, and cutting 107 more 
grooves with a feed rate of 0.8~mm/sec. 
The grooves had a designed pitch $p= 350$~\micron\ and designed depth $h=1000$~\micron\ to 
produce square pyramids with $w=52$~\micron. 
During machining, less than 1~\% of the pyramids broke. 
The total machining time was 3 hours. 

\section{Shape Measurements and Results }
\label{sec:measurements}

\subsection{Measurements}

\begin{figure}[h!] 
   \centering
   \includegraphics[width=8.1cm]{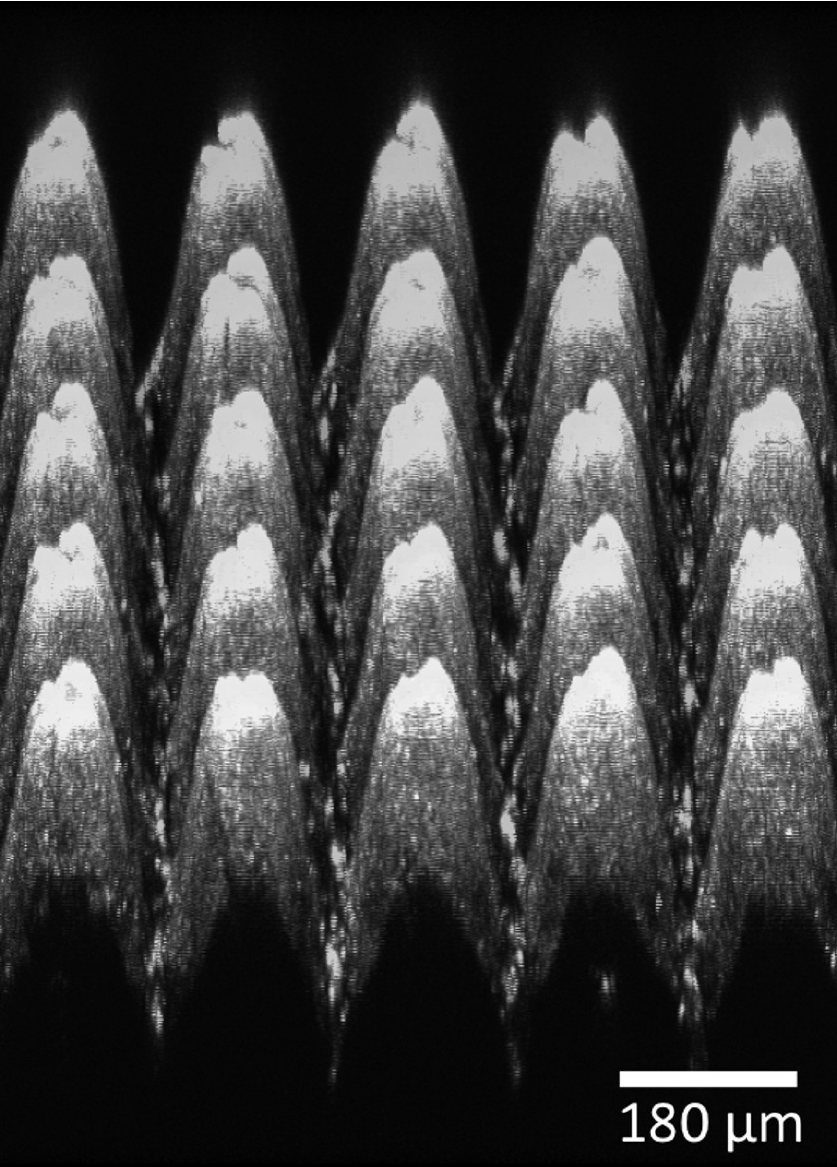} 
   \caption{ Perspective view of laser ablated silicon, sample \Lone . 
            The brightest parts are the peaks of the \ac{SWS}.
   \label{fig:si21_projection} }
\end{figure}

\begin{figure}[h!] 
   \centering
   \includegraphics[width=8.5cm]{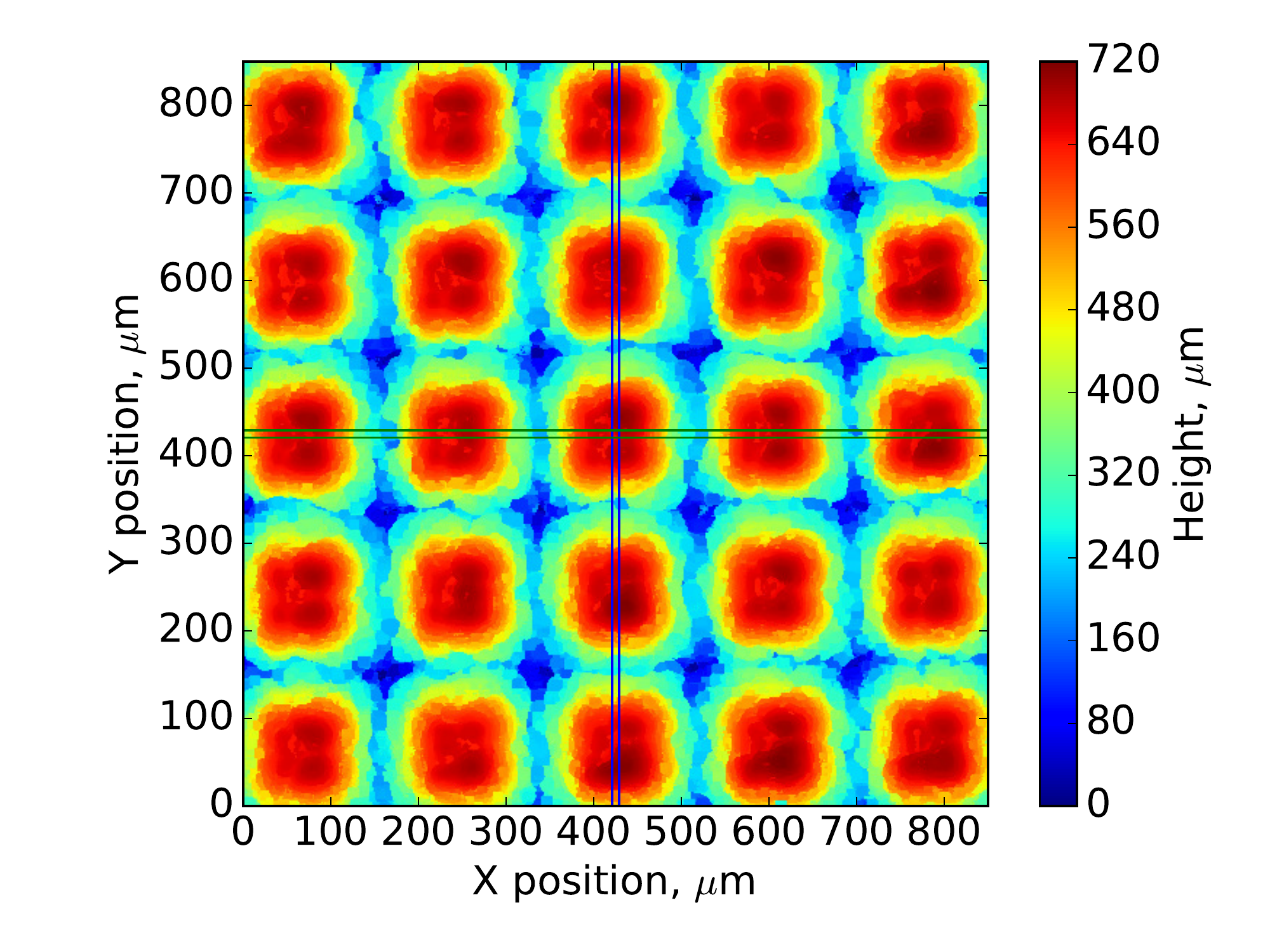} 
   \caption{ Height map of laser ablated silicon, sample \Lone. 
   Double blue and green lines mark the position along which we extracted the 1-dimensional height 
   profiles shown in Figure~\ref{fig:si21_cross}. The width between the lines corresponds to the portion of 
   the height map averaged to produce the profile.
   \label{fig:si21_shape} }
\end{figure}

We used a Nikon A1RMP confocal microscope to image \Lone\ and \Ltwo . 
The surfaces of \D\ were too smooth to produce sufficient diffuse reflection, so we 
used a Keyence VHX-5000 optical microscope.  
Both microscopes had $x,y$ plane resolution of 2~\micron. We imaged 9 locations 
on \Lone\ and 5 locations on each side of \Ltwo. At each location we took a 
series of images spaced in $z$ by 5~$\mu$m and constructed a
3-dimensional image and a height map. Examples are shown in  
Figures~\ref{fig:si21_projection} and~\ref{fig:si21_shape}, respectively. 

Using the images, we measured the geometrical properties of the samples, including the height $h$, pitch $p$, 
and, where relevant, the width of the peak $w$. The ablation caused deeper troughs at the intersection of grooves.  
We determined the height of the \Lone\ and \Ltwo\ pyramids by finding the median $z$ coordinate of the peaks 
and troughs, the red and blue regions seen in Figure~\ref{fig:si21_shape}.
The measured height is quoted as the difference between these medians.  Table~\ref{tab:geo} 
gives the mean height, pitch, and width, and the standard deviations of the values across all imaged areas. 

\begin{table}[h]
   \caption{Geometric parameters of SWS. 
   \label{tab:geo}  }
   \begin{tabular}{c|c|c|c}
	  \hline \hline
	  Sample  & Height$^a$   & Pitch$^a$  & Peak width$^a$ \\ 
                  &  [$\mu$m] & [$\mu$m] & [$\mu$m]     \\ \hline             
          \D      & $972 \pm 3$ &  $350 \pm 2$  & $51 \pm 3$ \\ %\hline
          \Lone   & $720 \pm 20$ &  $182 \pm 2$  & N.A.$^{b}$ \\
          \Ltwo  (side A)   & $600 \pm 15$ &  $179 \pm 4$  & N.A.$^{b}$ \\
          \Ltwo  (side B)   & $560 \pm 20$ &  $179 \pm 3$  & N.A.$^{b}$ \\
          \hline \hline
          \multicolumn{4}{l}{\footnotesize$^a$The uncertainty quoted is the standard deviation} \\ %\hline
          \multicolumn{4}{l}{\footnotesize from multiple imaged areas.} \\ %\hline
          \multicolumn{4}{l}{\footnotesize$^b$\Lone\ and \Ltwo\ did not have well-defined flat peaks.} %\\ %\hline
   \end{tabular}
\end{table}

\begin{figure}[h!] 
   \centering
   \includegraphics[width=8.5cm]{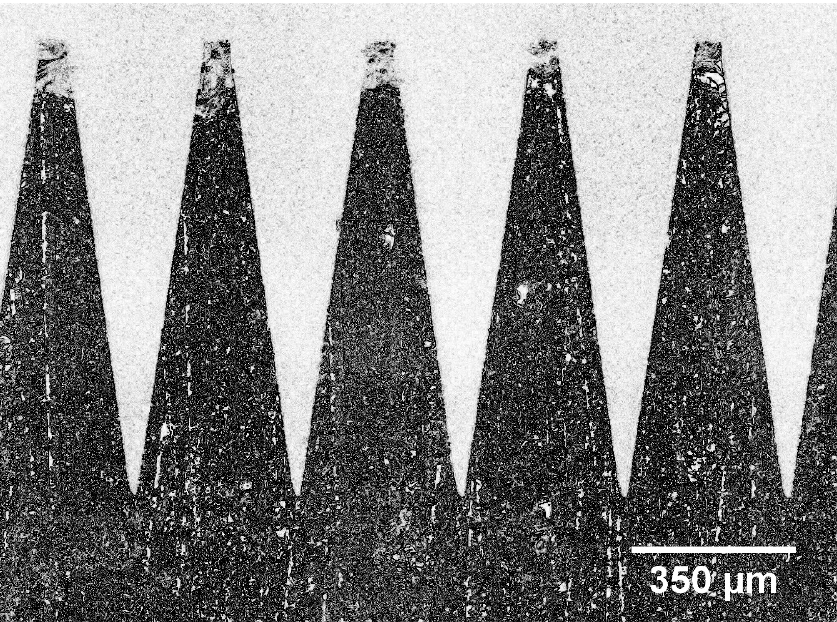} 
   \caption{ Side-view image of \D. The dark areas are silicon and the light are air.
   \label{fig:si01_shape} }
\end{figure}

For \D\ we found it more instructive to image a side view along 
the $y$-axis and view the \ac{SWS} in profile in the $x-z$ plane.  
This view is shown in Figure~\ref{fig:si01_shape}. For this sample, 
since the grooves were of uniform depth, the pyramid height was the same as the depth of the groove. 
We measured the depth of 17 grooves in the center of the sample by imaging the cut profiles. 
The mean and standard deviation are given in Table~\ref{tab:geo}.  
To characterize blade wear we also imaged two grooves, one made 100 cuts after the other, a total 
of cut length 465~cm. The depth difference was 33~\micron. 

\subsection{Comments Regarding the Laser Ablated Shapes}

The measured pitch of \Lone\ and \Ltwo\ matches the design value. It is a factor of 2.2 smaller than 
the pitch reported by \textcite{datta2013} who
used sequential dicing with commercially available dicing saws. It is a factor of 1.8 smaller than the 
pitch of the \ac{SWS} we recently laser-ablated on alumina and sapphire.\cite{matsumura2016_arc}  

The height of the \Ltwo\ pyramids was 17~\% smaller than the height of pyramids on \Lone, while the 
laser power was only 5~\% lower. We suggest the following explanation. \Ltwo\ was three times thicker
than \Lone, implying that during ablation at similar incident power \Lone\ was likely to be hotter than \Ltwo. 
\textcite{thorstensen2012} reported a decrease in the ablation threshold of silicon 
with increasing temperature. Therefore the ablation threshold of \Lone\ was lower than that of \Ltwo. 
In an earlier publication~\cite{schutz2016} we noted that ablation along a sloped face 
stops when the energy density of the beam dilutes below the ablation threshold. 
The effect is purely geometrical---%
a normally incident beam becomes elliptical when projected onto the ablated, sloped surface---%  
and thus the maximum slope is calculable from a known ablation threshold. 
The higher the ablation threshold, the shallower the maximum slope. 
We hypothesize the combination of lower laser power and higher ablation threshold, 
due to the cooler substrate, produced 
a shallower slope angle on \Ltwo\ of 84\degree\ vs 85\degree\ on \Lone.  This is sufficient to 
explain the difference in heights. 

\begin{figure}[h!] 
   \centering
   \includegraphics[width=8.5cm]{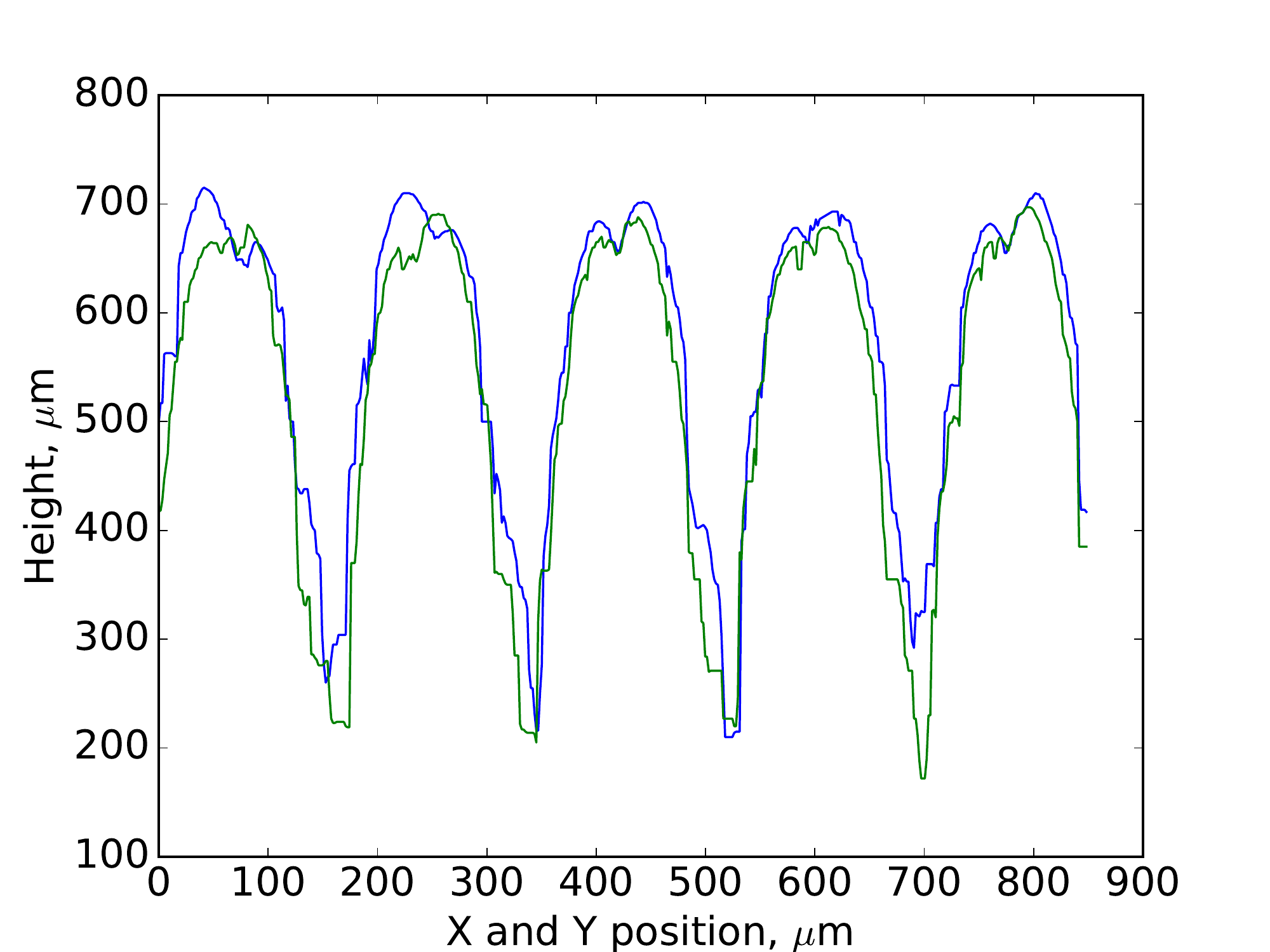} 
   \caption{ Height profiles of \Lone. The profiles were produced from cuts through peak centers, as shown in 
           Figure~\ref{fig:si21_shape}. Cuts at constant $y$ are in green; constant $x$ are in blue.
   \label{fig:si21_cross} }
\end{figure}

The ablation images show most of the pyramid tips are cracked. This may be a consequence of 
excessively high energy density. With this silicon ablation we are using 
$1.4 \cdot 10^{6}$ J/m$^{2}$/pulse. With sapphire, also a single crystal, we used an energy 
density per pulse that was 15 times lower and observed no cracking. The majority of the 
factor of 15 is due to the 3.75 times smaller spot diameter we used with silicon, a consequence of striving 
to reach smaller pitch and pyramid tips.  
The ablated shape repeatability is good as shown in Figure~\ref{fig:si21_cross}. 

\subsection{Comments Regarding the Diced Pyramid Shapes}

The SWS on \D\ were a regular array of truncated, square-based pyramids which matched the designed 
pitch of 350~\micron\ but were 25~\micron\ shorter than the 1000~\micron\ designed height.  The height discrepancy 
was likely due to uncertainty in the absolute surface position of the wafer while being machined.
However, the cut-to-cut repeatability in height was 3~\micron\ or better as shown by the small measured 
variation in height across 17 successive cuts. 

We measured a 33~\micron\ decrease in structure height after 100~cuts, 465~cm of cut length. 
This difference was due to either blade wear or a slightly tilted sample mount.
Assuming the entire height difference was due to blade wear, we calculated 0.07~\micron\ of 
wear per cm of cut length.
If a large sample was machined, it would be important to re-zero the blade position or continuously 
adjust the depth to account for wear. Measurements on a different blade indicated negligible 
wear over 160~cuts, 720~cm of cut length. More data is needed before definitive information 
is available regarding beveled blade wear. 

Sample \D\ showed a negligible number of broken pyramids. We did not attempt to optimize the trade-off between 
structure robustness, blade feed-rate, and blade wear.  \textcite{datta2013} used a factor 
of 50 higher feed-rate, suggesting the feed-rate with the beveled saw can be increased.

\section{Optical Performance}
\label{sec:reflect}

\subsection{Measurement Procedure}
\label{sec:measurementprocedure}

We measured the reflectance of \Lone, \Ltwo, \D, and \Ftwo\ in two polarizations and six
frequency bands between 70 and 720~GHz: 70--120~GHz, 110--170~GHz, 140--260~GHz, 
215--320~GHz, 310--480~GHz, and 460--720~GHz. 
We measured \Fone\ at the same frequencies, but in one polarization only. 
We measured transmittance of \Ftwo, \Lone, and \Ltwo\ in four bands between 70 and 320~GHz: 
70--120~GHz, 110--170~GHz, 140--260~GHz, and 215--320~GHz.  
The measurements were made at the Institute for Terahertz Science and Technology 
(ITST) at the University of California, Santa Barbara.  The reflectance setup is described in \textcite{bailey2015} 
The transmittance setup was similar but the sample was placed 
just after the source, and a gold mirror replaced the reflectance sample. 
All measurements with samples were normalized using data runs without samples. 

\begin{figure}[h] 
   \centering
   \includegraphics[width=8.5cm]{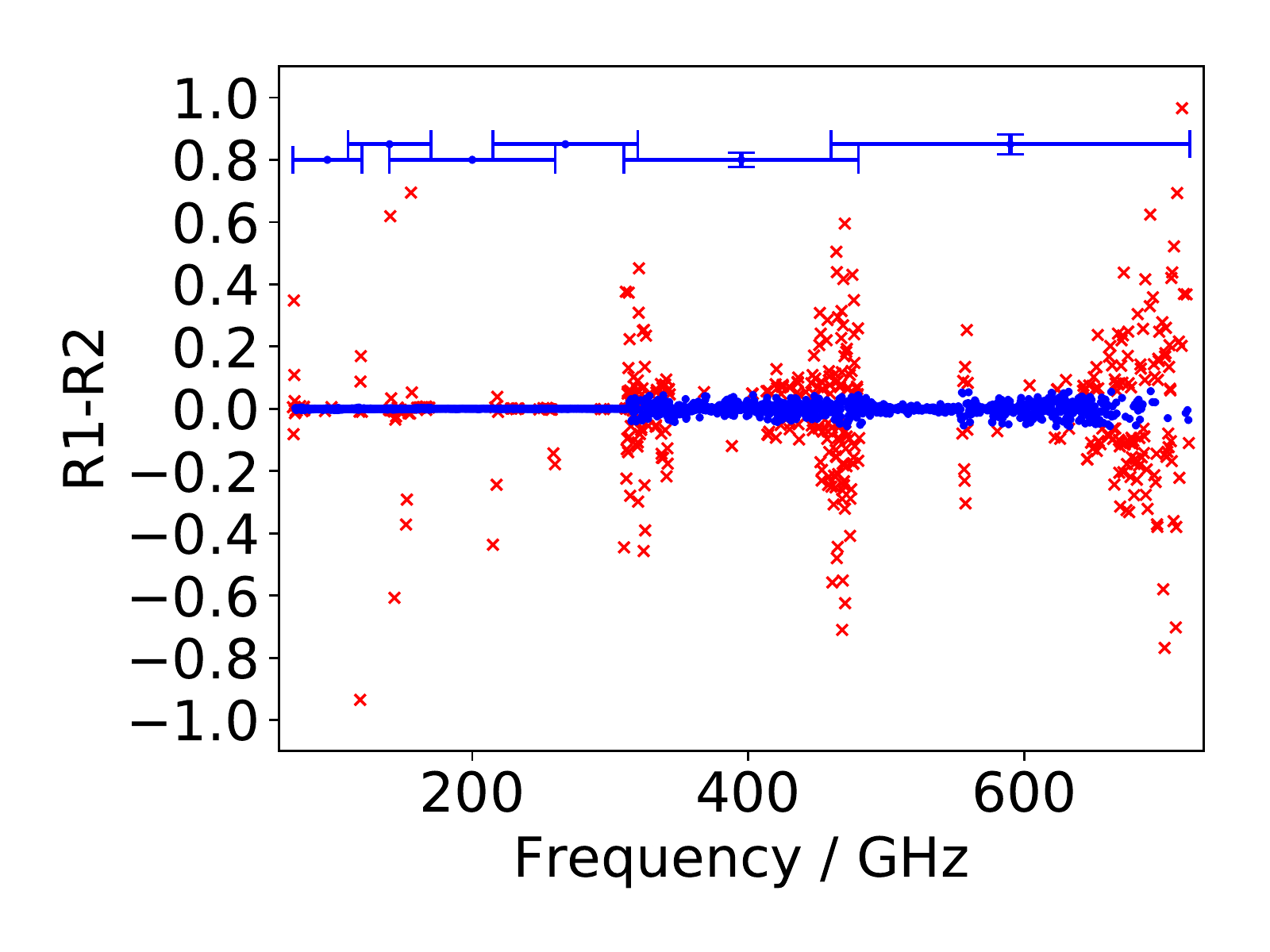} 
   \caption{Difference of two successive reflectance measurements of sample \Fone\ including data 
            that pass (blue dots) and are rejected (red crosses) by the 4.5 median absolute 
            deviation (MAD) criterion. Estimated measurement error per point (blue)
            is shown with one representative error per band at an arbitrary vertical offset 
            for clarity. The horizontal extent is the bandwidth. See the text and the Appendix for more details.  
   \label{fig:ucsb_data_diff} }
\end{figure}

The ITST data were taken twice for each measurement without changing the setup.  
The difference between these measurements was used to remove outliers and estimate 
a measurement error per point. An example of one such difference, from the 
pair of measurements on \Fone , is shown in Figure~\ref{fig:ucsb_data_diff}. 
There are two primary sources of higher levels of noise: (1) lower source power output
near band edges, and (2) an atmospheric water line at 557~GHz.\cite{vanExter89}
From the pair difference data we calculated the mean and the median absolute 
deviation (MAD) per band. The means were always within one MAD of zero. 
We removed as outliers all data with difference greater than $4.5$~MAD 
from the mean. We calculated the distribution and cumulative distribution for the pair
difference data and assigned the measurement uncertainty per point 
as the level at which the cumulative distribution reached 68~\%. These measurement 
uncertainties are shown for all data sets as single error bars per band. For 
clarity, they are displayed at an arbitrary 
vertical offset. The horizontal extent is the bandwidth over which the measurement 
error was calculated.  
A more detailed description of the data cuts and measurement uncertainty estimation 
is given in the Appendix.

\subsection{Measurements}

\begin{figure}[h] 
   \centering
   \includegraphics[width=8.5cm]{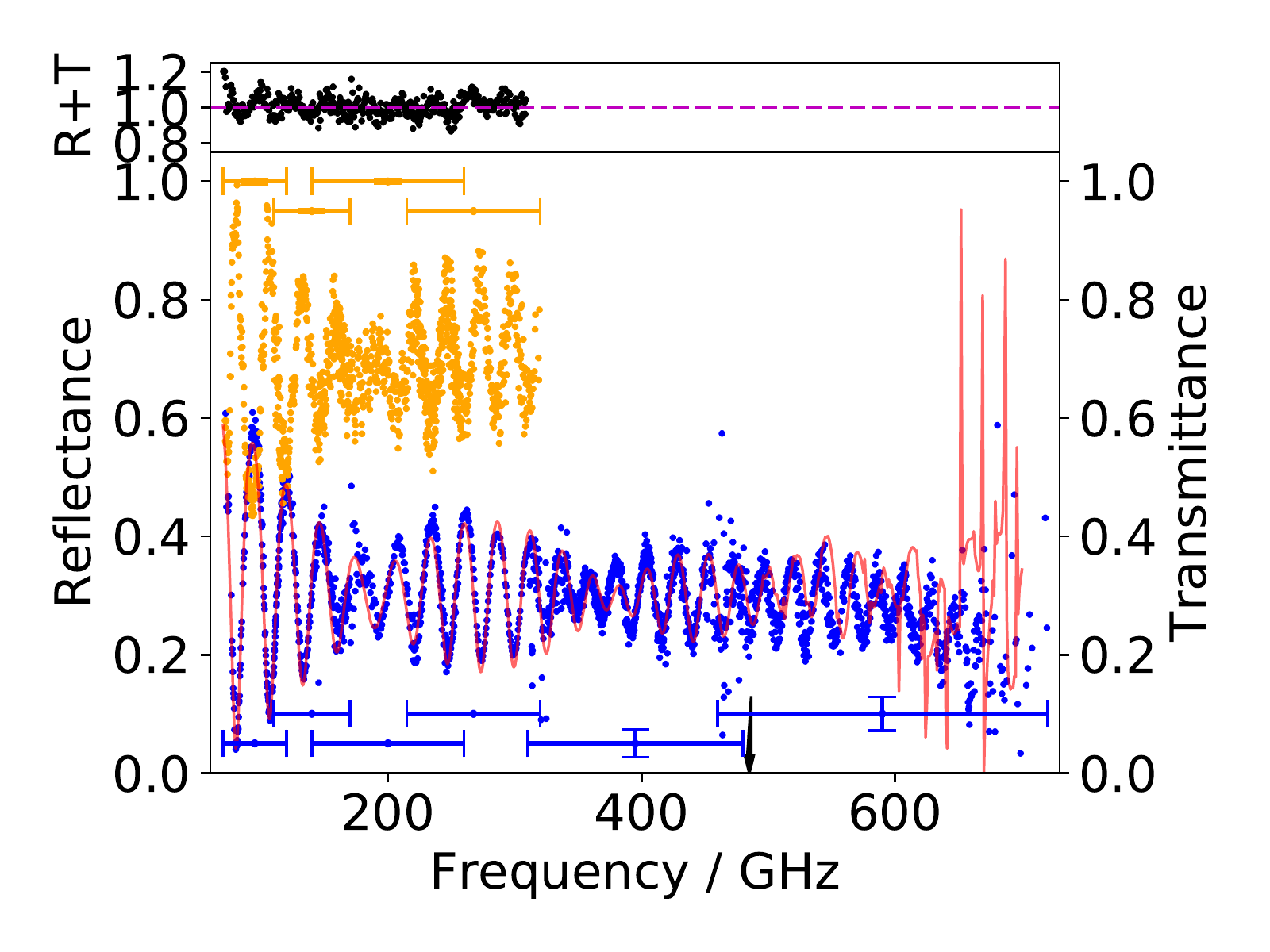} 
   \caption{ Reflectance (blue), transmittance (orange), and their sum (black) of \Lone\ as a function of frequency. 
   The average reflectance is near 0.3 because this sample was laser ablated on one side only. 
   \ac{FEA} predictions (red) match the data well up to frequencies somewhat above 
   $\nu_d = c/(n p) = 485$~GHz (black arrow), 
   beyond which diffraction is expected to be significant; see Section~\ref{sec:sim}. 
    Measurement errors per data point (orange and blue) are adjacent to their respective data and are discussed
    in Section~\ref{sec:measurementprocedure} and Figure~\ref{fig:ucsb_data_diff}.
   \label{fig:si21_ucsb} }
\end{figure}

\begin{figure*}[ht] 
   \centering
   \includegraphics[width=17cm]{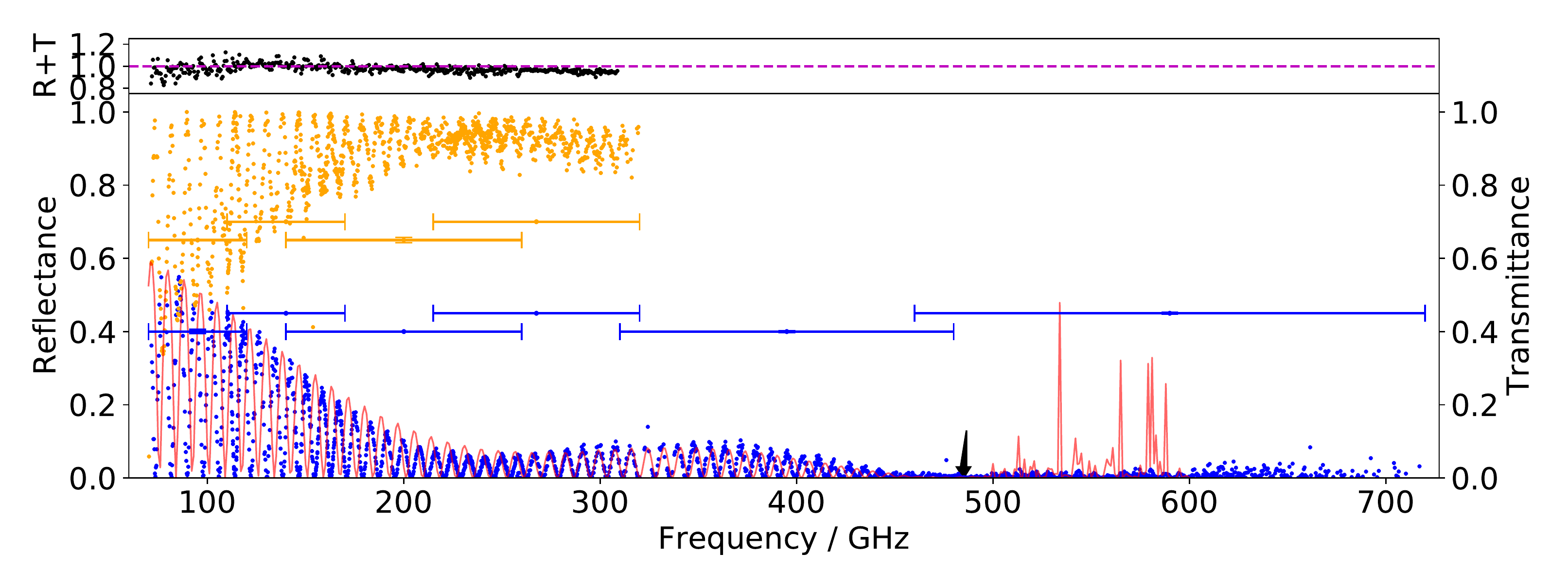} 
   \caption{Reflectance (blue), transmittance (orange), and their sum (black) of \Ltwo\ as a function of frequency. 
   \ac{FEA} predictions (red) match the overall reflectance envelope up to $\nu_d = c/(n p) = 490$~GHz (black arrow), 
   beyond which diffraction is expected to be significant (see Section~\ref{sec:sim}) and is observed in the FEA data.  
   \label{fig:si22_ucsb} }
\end{figure*}

\begin{figure}[h] 
   \centering
   \includegraphics[width=8.5cm]{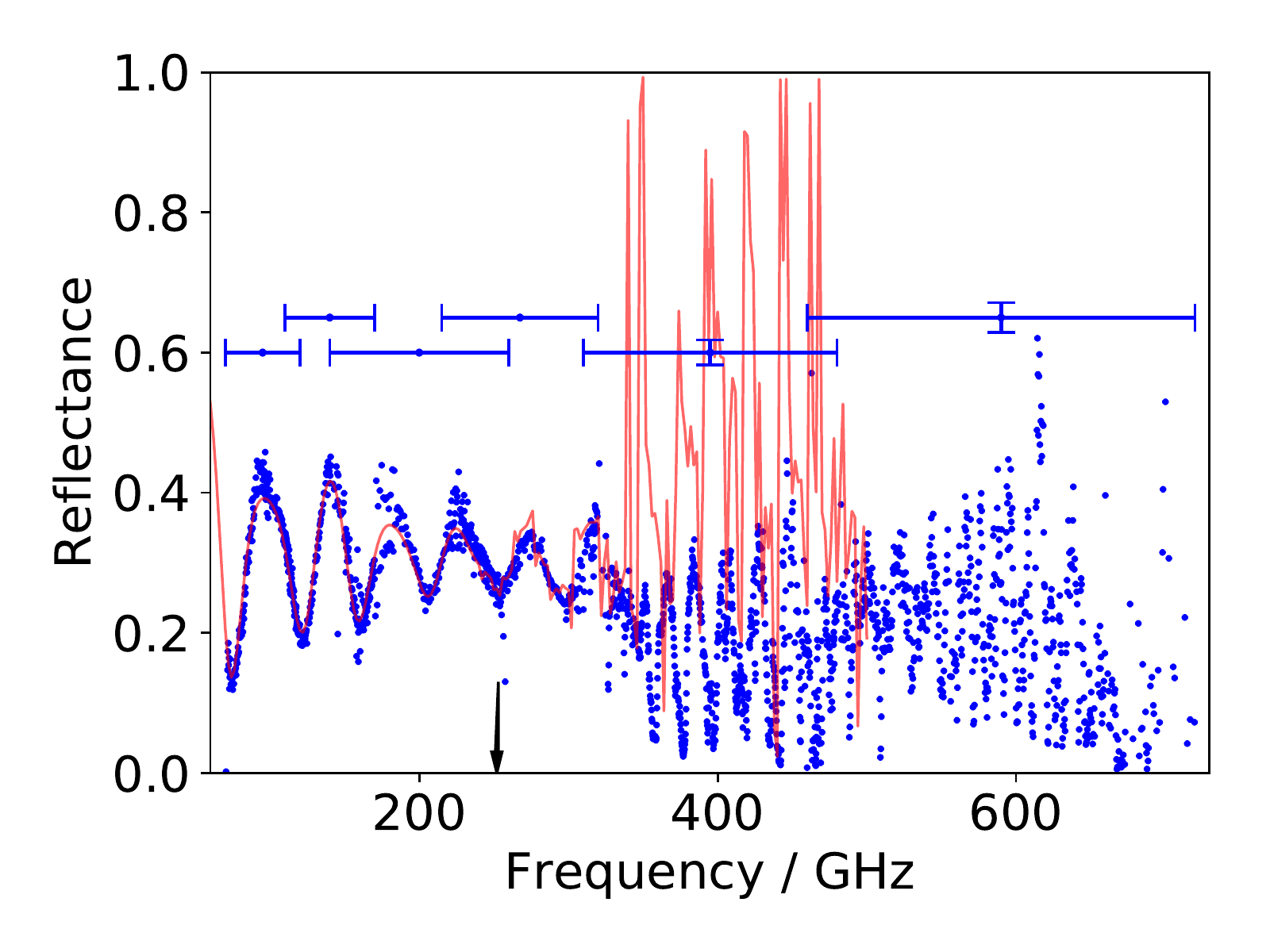} 
   \caption{Reflectance of \D\ (blue) and FEA simulations (red) using the geometry shown in  
            Table~\ref{tab:geo}.  The black arrow at $\nu_d =  c/(n p)
            = 252$ GHz is where diffraction first becomes significant within the sample. 
   \label{fig:si01_ucsb} }
\end{figure}

Figures~\ref{fig:si21_ucsb}, \ref{fig:si22_ucsb}, and \ref{fig:si01_ucsb} show reflectance 
measurements of \Lone, \Ltwo, and \D\ as well as the sum of transmittance and
reflectance, where relevant.  The Figures also 
give the upper frequency limit $\nu_d = c/(n p)$ where diffraction becomes significant 
due to the pitch of the \ac{SWS}. We discuss this diffraction in Section~\ref{sec:sim}.  

\begin{figure}[h] 
   \centering
   \includegraphics[width=8.5cm]{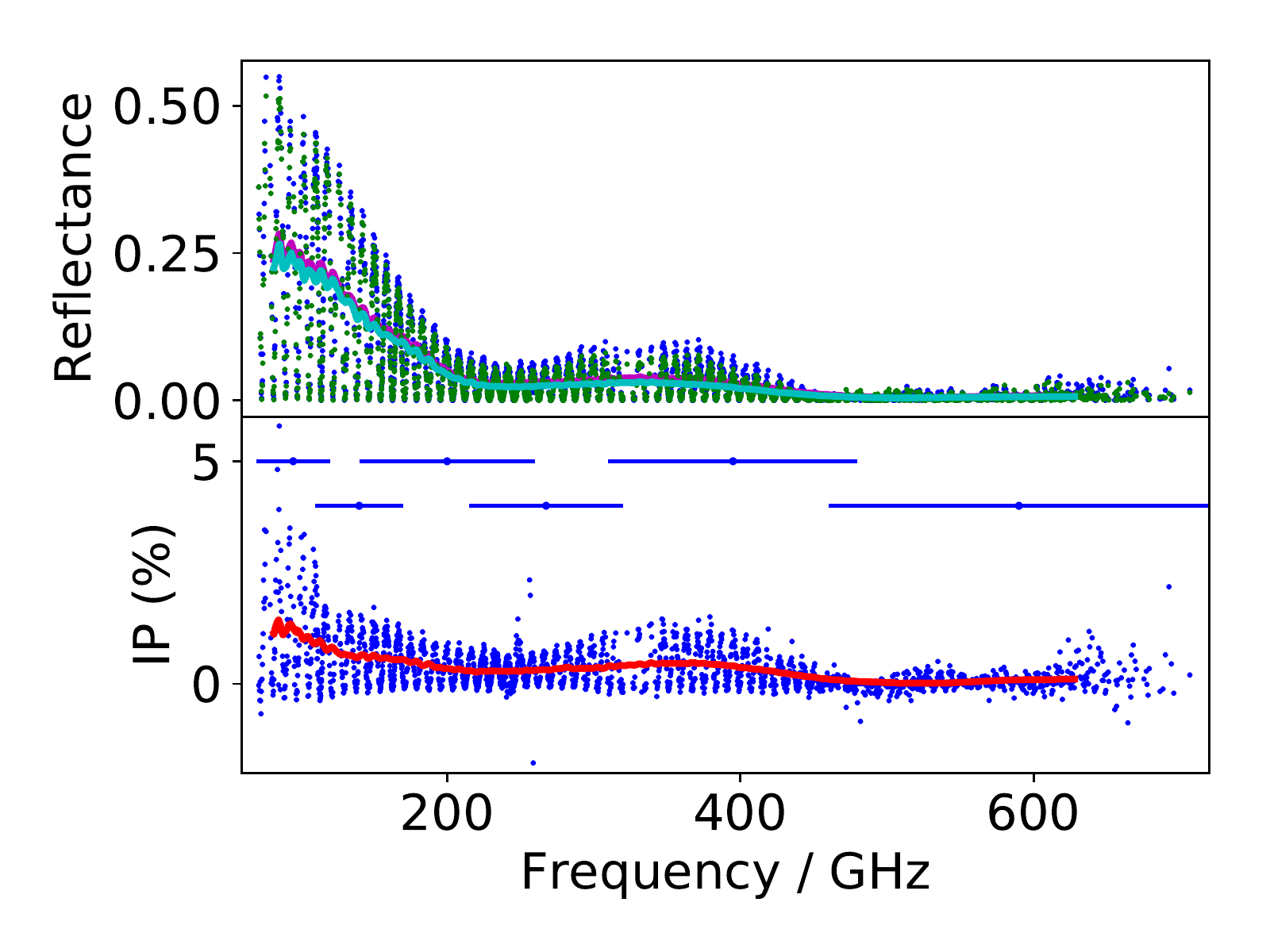} 
   \caption{Upper panel: Reflectance from two orthogonal polarizations (blue, green) for \Ltwo , and 
   running averages over a 25~\% fractional bandwidth (magenta and cyan). 
   Lower panel:  Instrumental polarization 
   calculated from the reflectance data (blue) and a running average over 25~\% fractional bandwidth (red). 
   \label{fig:pol} }
\end{figure}

To characterize the polarization properties of the samples we calculated the difference in reflectance between 
measurements in two orthogonal polarizations.  Figure~\ref{fig:pol} shows the reflectance data for \Ltwo. It also 
shows the level of instrumental polarization (IP) defined as 
\begin{eqnarray}
IP = \frac{T_{\parallel} - T_{\perp}}{T_{\parallel} + T_{\perp}}. 
\end{eqnarray}
Instrumental polarization represents the level of conversion of unpolarized to polarized light by an instrument 
or one of its components. 
To calculate IP we used the reflectance data for each sample and assumed low loss; $T = 1 - R$.
For \Lone, \Ltwo, and \D\ the maximum IP, averaged across 25~\% fractional bandwidth, from 
70--700~GHz was 1.2~\%, 1.4~\%, and 1.6~\% respectively. 

Measurements in two polarization states on \Ftwo, for which the IP should have been zero, 
gave band averaged (${\Delta \nu} / {\nu} = 25~\%$) IP of less than $0.6~\%$ below 320~GHz, 
less than $0.7~\%$ from 320--490~GHz, and less than $1.5~\%$ from 490--700~GHz. 
This indicated that below 490~GHz the measured IP due to the \ac{SWS} was at or below the 
$\sim 1$~\% level. 
Above 490~GHz we placed an upper limit of 1.5~\% on the IP.

\subsection{Modeling}
\label{sec:sim}

We modeled transmittance and reflectance using an electromagnetic finite element analysis code (HFSS)\cite{hfss}
and in some cases augmented it with calculations using rigorous coupled-wave 
analysis (RCWA).\cite{moharam1995,rcwa_software}

We used  the 3-dimensional image of \Lone\ to construct a solid model of a single pyramid. 
This model pyramid was imported into the FEA and placed on a 1.287~mm thick substrate 
(the thickness of the native sample minus 0.720~mm, the height of \Lone). An infinite planar sample was  
simulated using periodic boundary conditions.  

We reconstructed a single pyramid for each side of \Ltwo, placed them on each side of a 4.871~mm substrate, 
and simulated \Ltwo\ in the same way as \Lone. To facilitate periodic boundary conditions we ignored 
the 45\degree\ rotation between the patterns on the two sides. 

For \D, we constructed a square based, truncated 
pyramid using $h$, $p$, and $w$ as measured and reported in Table~\ref{tab:geo}. The pyramid was 
placed on a 1.037~mm thick substrate and duplicated using periodic boundary conditions.  

The results of the FEA simulations are shown in 
Figures~\ref{fig:si21_ucsb}, \ref{fig:si22_ucsb}, and~\ref{fig:si01_ucsb}.  There was generally 
good agreement between simulations and data at frequencies up to and somewhat above 
the `diffraction frequency' $\nu_{d} = c/(n p)$, which is marked with a black arrow. For \Ltwo\ 
the data have a fringe pattern that has 4~\% lower frequency than the simulation. We comment
on this in Section~\ref{sec:perform}. The frequency $\nu_{d}$ was the lowest frequency at which we expected 
diffraction and constructive interference inside the substrate to be significant. The constructive interference 
was due to the periodic structure of the pyramids and occurred at frequencies above
\begin{eqnarray}
  \nu_{d,lm}                     = \frac{c}{np} \sqrt{l^2 +  m^2}.
\end{eqnarray}
where $l$ and $m$ are integers. At frequencies above $\nu_{d} = c/(n p)$ the FEA showed 
strong reflections, but the data do not. 

\begin{figure}[h]%tbp] %  figure placement: here, top, bottom, or page
   \centering
   \includegraphics[width=8.5cm]{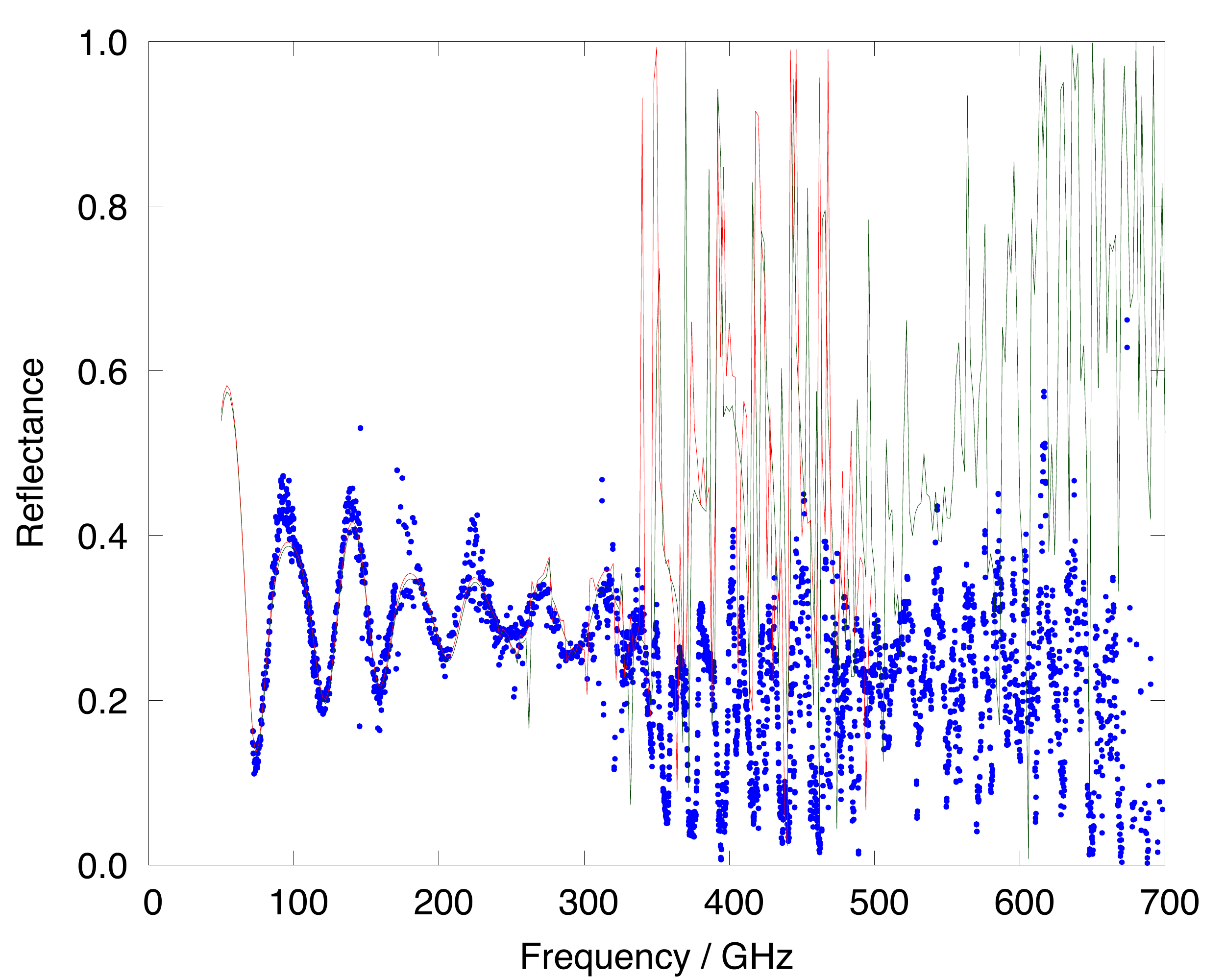} 
   \caption{Top: The measured reflectance (blue), and the simulated reflectance using RCWA (green) and HFSS(red).
           }
   \label{fig:reflectance.pdf}
\end{figure}

We cross-checked the results of the FEA using RCWA. 
Figure~\ref{fig:reflectance.pdf} shows the measured reflectance of \D\ 
and the simulated reflectance 
using RCWA and FEA. All three were consistent below $\nu_d=252$~GHz.
Above 252~GHz, at which the effect of diffraction was expected to appear with modes $(l,m)=(1,0)=(0,1)$,
RCWA and HFSS agree qualitatively; both showed high reflectance at similar 
frequencies, but the exact levels
were different. We propose the following explanation for the discrepancy between simulations and data.
The simulations duplicated one pyramid to create exactly periodic, infinite structures. Exactly periodic, 
infinite structures lead to high $Q$ resonances which produce sharp spikes in reflection.
However, the real sample was finite and consisted of pyramids which were all different in shape. 
A detailed comparison between the FEA, RCWA, and the data in the diffraction regime is beyond 
the scope of this paper, 
but it is interesting to note that measured data did not show the strong diffraction features 
suggested by simple 
calculations, indicating that the usable bandwidth of SWS-ARC may not be limited by $\nu_{d} = c/(n p)$.

\subsection{Comments Regarding Optical Performance}
\label{sec:perform}

\begin{figure}[h] 
   \centering
   \includegraphics[width=8.5cm]{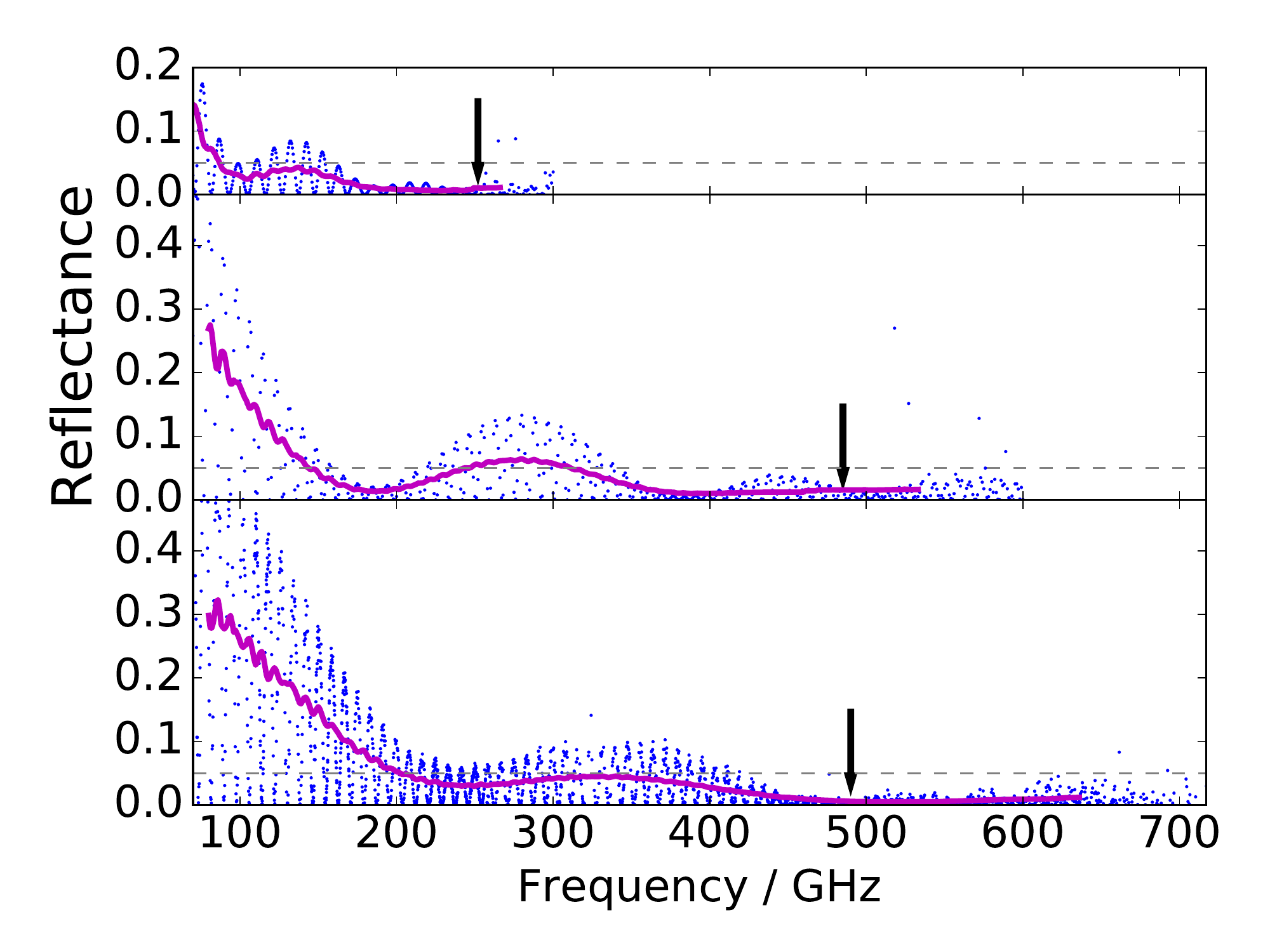} 
   \caption{Comparison of the three samples in reflectance (blue) and reflectance averaged over 25~\% 
            fractional bandwidth (magenta). 
            \D\ (top) and \Lone\ (middle) are simulations of two-sided samples while \Ltwo\ (bottom) is 
            measured data.  The dashed black lines indicate reflectance of 5~\%.
   \label{fig:2side} }
\end{figure}

Since FEA simulations agreed with measurements 
below $\nu_d$, we assessed performance of \Lone\ and \D\ up to $\nu_d$ by simulating silicon substrates coated on 
both sides with \ac{SWS} matching those measured for \Lone\ and \D.  These simulations are shown in 
Figure~\ref{fig:2side} along with the measured reflectance of \Ltwo. The Figure also shows reflectance averaged
over 25~\% fractional bandwidth, as would be appropriate for typical CMB experiments using bolometers. 

Beveled dicing saws are an efficient way to remove material and are thus suitable for fabricating relatively 
deep, large pitch structures. Therefore, beveled saws are particularly suitable for applications  
at lower frequencies. Averaged over ${\Delta \nu}/{\nu} = 25~\%$ fractional bandwidth,
reflectance drops below 5~\% at 87~GHz. Doubling of the height of the pyramid would halve this frequency 
to approximately 44~GHz.  Laser ablation is more efficient when patterning smaller structures. 
When averaged with 25~\% fractional bandwidth, reflectance on the simulated two-sided \Lone\ 
drops below 5~\% at 144~GHz.  The bandwidth with less than 5~\% reflectance
extends up to $\nu_{d}=485$~GHz, and higher if $\nu_{d}$ turns out to not limit the performance
of the \ac{SWS}. That these are realistic predictions is demonstrated by measurements of \Ltwo\ (Figure~\ref{fig:2side}). 
Defining the high frequency edge of an `effective band' as $\nu_{d}$ and the low frequency edge 
as $\nu_{l}$, we list the effective bands for \Lone, \Ltwo, and \D, in Table~\ref{tab:perform}.
The frequency $\nu_{l}$ is that frequency at which 
the 25~\% fractional bandwidth averaged reflectance drops below 5~\%. The Table also gives
the average reflectance within the effective band. 

\begin{table}[h]
   \caption{Optical performance of SWS-ARC.}
   \label{tab:perform}  
   \begin{tabular}{l|c|c}
	  \hline \hline
	  Sample  & Effective band  &  Average reflectance$^{a}$   \\ 
                  &  [GHz] & [\%]       \\ \hline             
          \Lone$^b$   & 144--485  &  2.0~\%     \\
          \Ltwo       & 202--490  &  2.9~\%   \\
          \D$^b$      & 87--252  &  2.8~\%    \\ 
          \hline \hline
          \multicolumn{3}{l}{\footnotesize$^a$ Averaged over the Effective band.} \\ 
          \multicolumn{3}{l}{\footnotesize$^b$ From two sided FEA simulation using measured} \\ 
          \multicolumn{3}{l}{\footnotesize ~~~\ac{SWS} shape.}  
   \end{tabular}
\end{table}

We attribute the difference in fringe frequency between the simulation of \Ltwo\ and data (Figure~\ref{fig:si22_ucsb}) 
to a crude approximation of the sample. As discussed in Section~\ref{sec:sim}, the simulation is based on duplicating 
only one pyramid for each side of the surface, thus ignoring variations in shape, pyramid height, and thickness of the 
remaining material. 

All three data sets, the simulated two-sided \D\ and \Lone, and the measured \Ltwo, suggest 
a straight pyramid is not the optimal shape. There is a lobe of increased reflectance at frequencies
above the onset of low-reflectance. Impedance matching theory suggests the implementation of a 
Klopfenstein profile~\cite{grann1995} to reduce this lobe. 

\section{Summary}
\label{sec:summary} 

We presented two novel approaches to producing \ac{ARC} on silicon for the \ac{MSM}: laser ablation and 
dicing with beveled saws.  The optical performance is summarized in Table~\ref{tab:perform}.  The \ac{SWS}, laser 
ablated on both sides of a silicon disc, reduced 25~\% band averaged reflectance to below 5~\% over a fractional 
bandwidth of 83~\% centered on 346~GHz.  The dicing saw structures were larger, in 
both pitch and height, making them effective at lower frequencies. The dicing saw structures reduced 
reflectance to below 5~\% over a fractional bandwidth of 97~\% centered on 170~GHz. 
These effective bandwidths are sufficient for most currently deployed cosmic microwave background instruments, 
but would need to be increased to 150~\% fractional bandwidth to meet the needs of all upcoming instruments.
The effective band of both methods was limited by the structure height at low frequencies and pitch at 
high frequencies.

Machining times   
for both methods were similar. It took 3.4~hours to ablate one side of a 20~cm$^2$ disc and 
3~hours to machine 16~cm$^2$ with dicing 
saws.  Machining time for both methods can be improved by optimizing the machining parameters.
To our knowledge, this is the first SWS-ARC for the \ac{MSM} produced using these techniques.

%\section*{Funding Information}

\begin{acknowledgments}
We thank two anonymous referees for detailed and helpful comments.
The authors acknowledge use of resources provided by the 
University of Minnesota Imaging Center (\url{http://uic.umn.edu}) and 
Minnesota Supercomputing Institute (\url{http://www.msi.umn.edu}). 
The research described in this paper used facilities of the Midwest 
Nano Infrastructure Corridor (MINIC), a part of the National 
Nanotechnology Coordinated Infrastructure (NNCI) program of the National 
Science Foundation. 
The transmittance and reflectance measurements were performed at the ITST 
Terahertz Facilities at UCSB, which have been upgraded under NSF Award No. DMR-1126894.
This work was partially supported by JSPS KAKENHI Grant Number 15H05441; the 
Mitsubishi foundation (grant number 24, JFY2015, in science and technology); 
the World Premier International Research Center Initiative (WPI), MEXT, Japan; 
and the JSPS Core-to-Core Program, Advanced Research Networks.
\end{acknowledgments}

\begin{acronym}
    \acro{ARC}{anti-reflection coatings}
    \acro{CMB}{cosmic microwave background}
    \acro{MSM}{millimeter and sub-millimeter}    
    \acro{SWS}{sub-wavelength structures}
    \acro{FEA}{finite element analysis}
\end{acronym}

\appendix*
\section{Data Cuts and Error Estimation}

To remove outliers and estimate uncertainty per point we used the difference between two 
subsequent data runs on the same sample with the setup unchanged.  The difference vs frequency 
for \F\ is shown in Figure~\ref{fig:ucsb_data_diff} and we use these data as an example. The same 
procedure was used with pair differences for all measurements.

\begin{figure*}[hbt] 
   \centering
   \includegraphics[width=15cm]{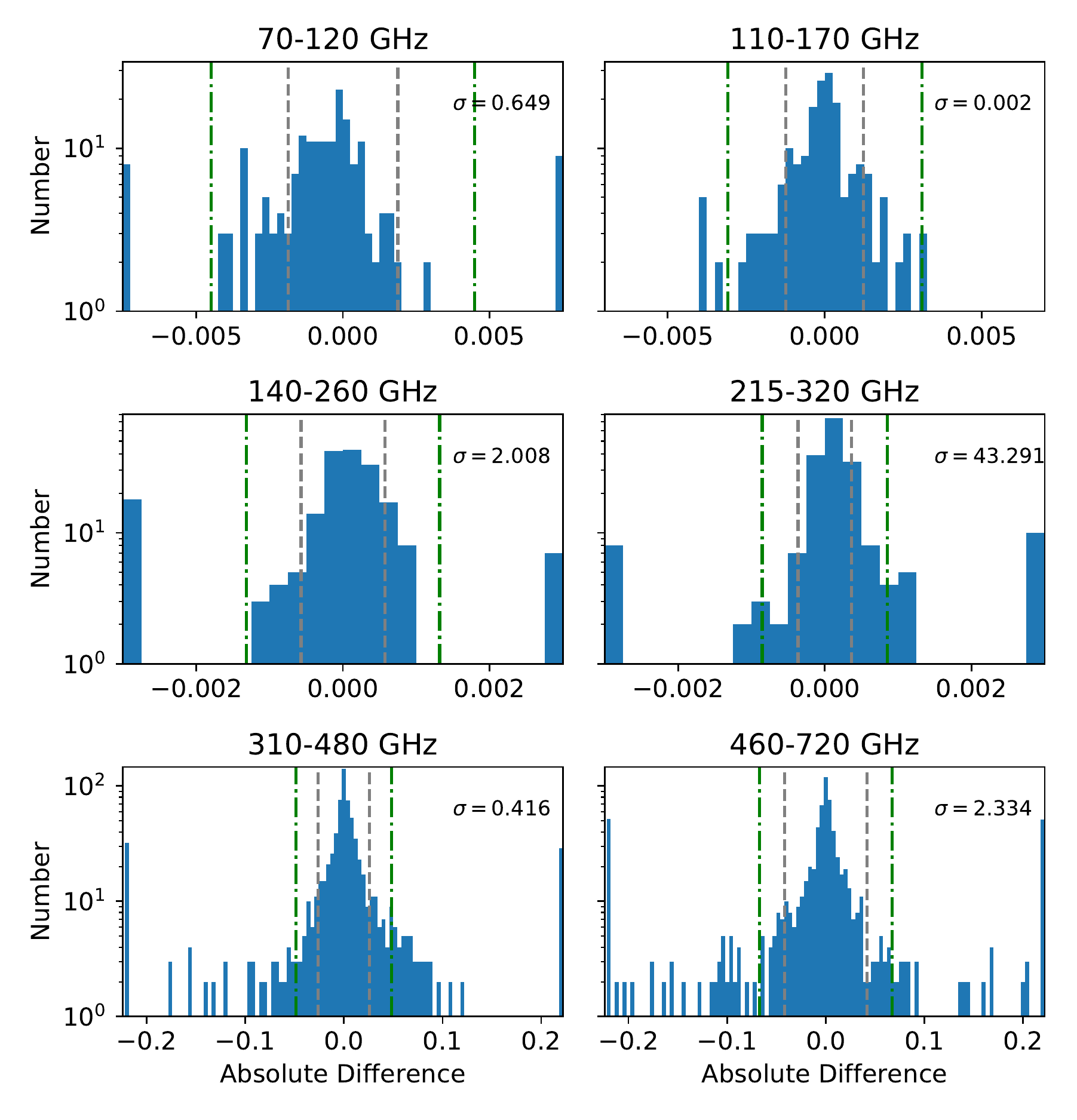} 
   \caption{Histograms of difference measurements of \F\, by measurement band.
            We only show the center of the distribution; note the varying ranges for the horizontal axes. 
            All values which fall outside the displayed range 
            are included in the leftmost and rightmost bins. For each panel we give the value of the standard deviation ($\sigma$), 
            which is typically outside the displayed region, as well as the  
            outlier criterion of 4.5~MAD (dot-dash) and the value of the estimated 
            error (dashed) lines.  
   \label{fig:histogram} }
\end{figure*}

\begin{figure*}[hbt] 
   \centering
   \includegraphics[width=15cm]{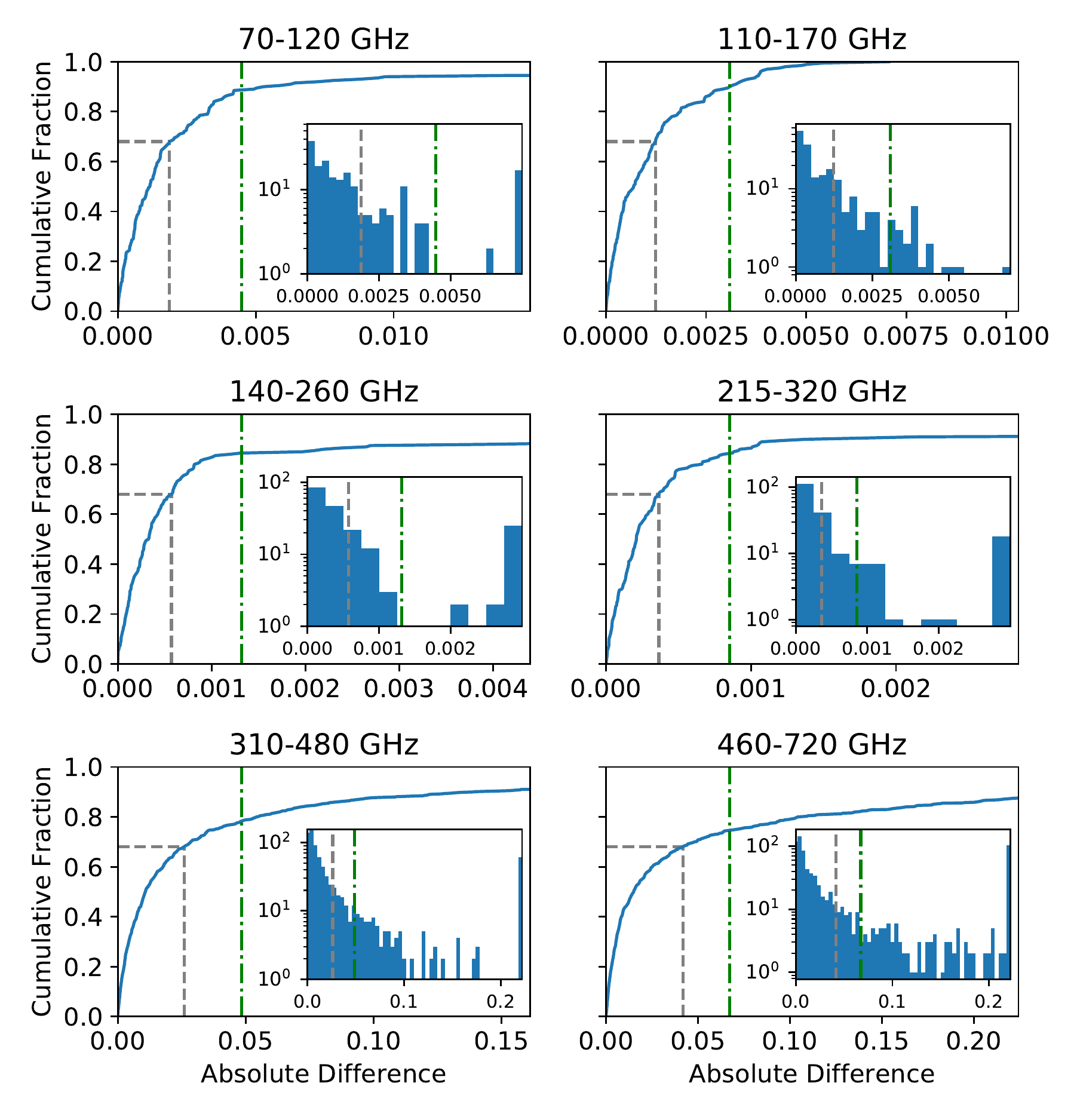} 
   \caption{Cumulative distributions and histograms (inset) of the absolute value of difference data per band with   
            the outlier criterion of 4.5~MAD (dot-dash) and determination of the estimated error using the 
            the 68~\% level of the distribution (dashed).
   \label{fig:cdf} }
\end{figure*}

Pair difference distributions and cumulative distributions for each of the 6 measurement bands are
shown in Figures~\ref{fig:histogram} and~\ref{fig:cdf}, respectively.  
The data are not Gaussian-distributed and therefore the standard deviation $\sigma$ is a poor 
estimator for the variance for the majority of the data. We 
used the median absolute deviation (MAD) instead.  For a Gaussian distribution
$\sigma / \text{MAD} = 1.48$, however this ratio ranged between 2.5 and $2\times 10^5$ for the 
6 pair difference measurements of \F.  The calculated standard deviations for each of the 6 bands 
are shown in Figure~\ref{fig:histogram}.

We removed all points that were more than 4.5~MAD from the mean of the difference data. 
This cut removed the majority of the long outlier tail;  see Figure~\ref{fig:cdf}. 
We chose this criterion because for a Gaussian distribution 
$3 \sigma = 4.5$~MAD, and because it was reasonable based on the cumulative distributions.  
We estimated measurement errors for each measurement and for each band by finding 
the difference value in the cumulative distribution of absolute values that included 68~\% of the raw 
data, that is, prior to outlier removal. This is shown by the grey dashed lines in Figure~\ref{fig:cdf}. 
We used the cumulative distribution of absolute values because the distribution of $|R_{1} - R_{2}|$ 
did not depend on which data set was labeled ${1}$ or ${2}$.
These measurement errors are plotted at an arbitrary vertical offset in 
Figures~\ref{fig:silicon_flat_ucsb}, \ref{fig:ucsb_data_diff}, \ref{fig:si21_ucsb}, \ref{fig:si22_ucsb}, \ref{fig:si01_ucsb}, and \ref{fig:pol}.
The horizontal extent shows the width of each of the frequency bands. 

We tested the sensitivity of our calculated values to various data cuts.  
The values which could be affected are average reflectance on \Ltwo,
and the maximum IP for \D, \Lone, and \Ltwo.
Using the cumulative distributions, we removed data in each band with the largest pair difference values using 
thresholds which kept 70~\%, 80~\%, 90~\%, or 95~\% of the data.  
Across these removal criteria we found average reflectance on \Ltwo\ changed by 0.1 percentage points.  
Maximum IP changed by less than 0.3 percentage points for \D\ and \Lone\ and by 0.8 for \Ltwo.

\bibliography{silicon_ARC}

\end{document}